# Influence of ion implantation on the magnetic and transport properties of manganite films.

M. Sirena[1], A. Zimmers[2], N. Haberkorn[1], E. Kaul[1], L. B. Steren[1,*], J. Lesueur[2], T; Wolf[2], Y. Le Gall[3], J.-J. Grob[3] and G. Faini[4].

[1] Instituto Balseiro – Univ. Nac. de Cuyo & CNEA, Av. Bustillo 9510, 8400 Bariloche, Rio Negro – Argentina.

[2] UPR5-LPEM-CNRS, Physique Quantique, E.S.P.C.I., 10 Rue Vauquelin, 75231 Paris, France.

[3] Institut d'Électronique du Solide et des Systèmes, UMR 7163, 23 rue du Loess – BP20, F-67037 Strasbourg Cedex 02, France.

[4] LPN-CNRS, Route de Nozay, 91460 Marcoussis, France

Abstract

We have used oxygen ions irradiation to generate controlled structural disorder in thin manganite films. Conductive atomic force microscopy CAFM), transport and magnetic measurements were performed to analyze the influence of the implantation process in the physical properties of the films. CAFM images show regions with different conductivity values, probably due to the random distribution of point defect or inhomogeneous changes of the local Mn3+/4+ ratio to reduce lattice strains of the irradiated areas. The transport and magnetic properties of these systems are interpreted in this context. Metal-insulator transition can be described in the frame of a percolative model. Disorder increases the distance between conducting regions, lowering the observed $T_{MI}$. Point defect disorder increases localization of the carriers due to increased disorder and locally enhanced strain field. Remarkably, even with the inhomogeneous nature of the samples, no sign of low field magnetoresistance was found. Point defect disorder decreases the system magnetization but doesn't seem to change the magnetic transition temperature. As a consequence, an important decoupling between the magnetic and the metal-insulator transition is found for ion irradiated films as opposed to the classical double exchange model scenario.

**I. Introduction.**

Since their rediscovery in the past decade [1], $A_{1-x}A'_{x}MnO3$ (A: La, A': Sr, Ba, Ca) manganites have attracted a lot of attention in the scientific community due to the possibility of studying many physical problems related to the strong correlation between their structural, transport and magnetic properties and their potential use in many technological applications. Very soon it became clear that disorder, the Mn-O bond distance and the Mn-O-Mn angle influenced the transport and magnetic properties of these materials [2,3]. Experimental studies on the effect of hydrostatic pressure and cations substitution on the magnetic order and metal–insulator transition of bulk compounds were also early reported [4,5]. In thin films, biaxial strains significantly modify the transport and magnetic properties of these systems [6-10]. The authors found that as the influence of biaxial strains was increased by reducing the film thickness, the transport carriers became localized, the magnetic transition temperature (Tc) was reduced [9] and the competence between ferromagnetism and antiferromagnetism in the system was increased [10].

Ion implantation has been considered a non thermal method that produces lattice disorder by introducing vacancy-interstitial pairs in a controlled way [11, 12]. Moreover, ion implantation has gained importance in recent years as a general method to develop nano-scale devices using not only manganite films [13-15] but also High Tc (HTc) superconductors [16, 17]. In this context, understanding the influence of point defects on the physical properties of manganite films has regain importance. Many efforts have been made to understand the transport properties of implanted manganite films using different ions, Ar [11], Ag [13], Cr [18], Fe [19] and others [14-15]. Despite this, very little can be found concerning the influence of these defects in the magnetic properties of manganite films. For instance, it has been generally assumed without measuring the magnetic properties that the resistivity peak, corresponding to the metal-insulation transition is close to the ferromagnetic transition [11]. However, some experimental results in these systems show a decoupling of Tc from the temperature corresponding to the metal-insulation transition ($T_{MI}$) [8,9]. Finally it should be noticed that most of the experimental results are expressed in terms of the implantation dose. Since the structural damage in these systems depends on several other factors, like ion energy, ion mass, even film thickness, etc. it is difficult to compare different experimental results.

The aim of this work is to present a systematic study of the influence of point defects disorder in the transport and magnetic properties of manganite films. We also present a simple method to quantify the structural damage in terms of a more "universal" magnitude than the irradiation dose: the number of defects per atom (dpa) of the structure. This magnitude better expresses the implantation damage done in the system and has proven to give excellent results as a parameter of the influence of point defect disorder in the physical properties of irradiated High Tc superconductor films [20].

**II. Experimental Details.**

$La_{0.75}Sr_{0.25}MnO3$ (LSMO) films with 50 nm thickness were grown on single crystal $SrTiO_3$ (100) substrates by DC magnetron sputtering from a stoichiometric ceramic target. The deposition was made in an Ar(90%)/O$_2$(10%) atmosphere at a total pressure of 400 mbar.

More details about the films fabrication method and characterization are given elsewhere [21]. The as-growth films were irradiated with oxygen ions (O$^+$) at 150 keV with doses $1\times10^{14}$ ion/cm$^2 \leq \phi \leq 10\times10^{14}$ ion/cm$^2$. The choice of the O$^+$ energy was made considering the authors previous experience with oxygen implantation in HTc [20]. With O$^+$ energy up to this value there is no specific chemical or physical effect expected when implanted in the oxide layer; it can be stopped by reasonable thickness of metal (around 200 nm) or photoresists (1 μm), which is important for developing different devices. We have chosen the film thickness such as having a rather constant damage profile across the film thickness when irradiated with this ion energy (figure 1). The estimated depth range is 220 nm and in these conditions the parameter quantifying the effect of ion irradiation in the system is expected to be the mean value of the depth damage distribution [20]. Considering this, the final dpa factor can be easily evaluated by dpa = $10^8$ $c$. $\phi$ / $d$. Where $c$ is the TRIM [22] calculated mean defect number per incident ion and per Å for these irradiation conditions and sample properties (i.e. film's thickness, Sr concentration, sample density and substrate), $\phi$ is the dose in ions per cm$^2$, and $d$ the material density in atoms per cm$^3$. The dpa calculated in this way is a universal estimation (doesn't depends of the ions energy, mass, films thickness, etc). It is a good parameter to quantify the structural damage done by ion irradiation. In our case, the damage done by the oxygen ions is around $3\times10^{-16}$ dpa/ion. The dpa for each ion species is given, at first order approximation, by the population of the ion in the unit cell.

Conductive atomic force microscopy (CAFM)[23] measurements were done in a Veeco Dimension 3100 ® SPM with a CAFM module. The scans were done using a diamond boron doped conductive tip in contact mode. Different probe polarizations and deflection set points were used to verify that the basic results don't depend on the measurements conditions. The CAFM results shown in the paper were obtained using a probe polarization voltage of 0.5 V and a deflection setpoint of 0.4V. The resistivity of the samples has been measured using a standard four probes configuration. The resistivity was measured as function of temperature from 4 K to 300 K and as a function of magnetic field between -5 T and 5 T. The temperature and field dependence of the magnetization have been studied using a Superconductor Quantum Interference Device (SQUID) magnetometer between 4 K and 400 K and ±5T respectively..

**III. Results.**

**III.1 Conductive atomic force microscopy**

A study centred in the influence of ion irradiation in the phase separation of manganite films using CAFM microscopy will be presented shortly [24]. However, to facilitate the discussion, the topography (left) and CAFM (right) 5 μm x 5 μm images of LSMO films irradiated with increasing dose (from top to bottom) are shown in figure 2. All the samples present conducting areas surrounded by insulating regions. As disorder increases, i.e. irradiation dose augments, the fraction of conducting surface reduces, i.e. the overall conductivity decreases. Vacancy-interstitials pairs locally break the Mn-O bonds, induce lattice strain enhancement [25] and $Mn^{3+}/Mn^{4+}$ ratio depletion [26]. All these effects tend to localize the carriers in the sample, reducing its conductivity. As the density of defects increases the distance between the conducting "islands" progressively increases. The origin of this phenomenon is probably related to the inhomogeneities of the point defect distribution or inhomogeneous changes of the local Mn3+/4+ ratio to reduce lattice strains of the irradiated areas. For higher irradiation dose the number of defects increases and the relative magnitude of fluctuations decrease reducing the number of conducting regions. It should be noticed that the length scale of the electronic inhomogeneities in the pristine sample is much smaller that the one observed in the irradiated samples. This result probably arises from the different origin of the electronic inhomogeneities in the different samples. The manganites films have an intrinsic cation disorder at the A-site of the crystalline structure and inhomogeneous strain fields due to the different cation radii of the La and Sr.

It should be noticed that the samples don't present two well defined characteristic conductivity values, as expected in a phase separated system [24]. Equivalent results were found by scanning tunnelling microscopy for $Pr_{0.68}Pb_{0.32}MnO_3$ single crystals [27] and polycrystalline $La_{0.8}Sr_{0.2}MnO_3$ films [28]. These results seem to indicate that irradiated manganites present, in a more precisely description, inhomogeneous electronic properties due to the inhomogeneous distribution of disorder and strain fields, instead of phase separation or phase competition. This distinction has already been done in a theoretical paper, based on calculations at the nano-scale by Shenoy and co-workers [29].

**III.2 Transport properties.**

Figure 3 presents the resistivity of the LSMO films irradiated with different doses as function of temperature. In general, as the disorder increases, the resistivity increases, the metal-insulator transition, $T_{MI}$, shifts to lower temperatures and carriers localization is progressively noticed at low temperatures, in agreement with previous works [11,14,18]. The film irradiated with the highest dose presents an insulator behaviour from room temperature to 100 K. This result indicates that the disorder limit so as to observe a metal-insulator transition in these structures is around 0.25 defects per atom For high temperatures ($T>T_{MI}$) the transport is driven by a thermal activated conduction mechanism, characterized by an energy gap, $\Delta$ (i.e. the energy difference between the mobility edge and the Fermi level) that increases linearly with increasing defect per atom (Figure 4). As disorder and strains increases, higher energies are required for the carriers to move. The effect of ion irradiation on the $\Delta$ of these systems is much higher (~100 meV) than the one found when only oxygen vacancies are created or removed (~31 meV) [25].

The reduction of the metal-insulator transition for increasing irradiation doses can be explained in the frame of a percolative model [30,31]. As the irradiation dose increases the distance between conducting regions increases (Figure 2) and lower temperatures are necessary for the conducting areas to grow, interconnect and give place to conduction paths and a macroscopic metal-insulation transition. This evolution as function of temperature was observed for polycrystalline manganite films [28]. It is worth noticing that both, $\Delta$ and $T_{MI}$ change linearly with the number of defects per atom and that the later parameter is independent of the fluency, the ions mass, energy and the film thickness. In other words, this behaviour depends only on the material and the dependence of the energy gap and $T_{MI}$ with the dpa becomes a good measure of the compound sensibility to point defect disorder.

In a recent letter, Moshnyaga and co-workers propose that the metal-insulator transition is characterized by the dimensionless constant, $\alpha_\rho = (d\rho/\rho)/(dT/T)$, which it is assumed to quantify the disorder in the sample [32]. We have found that for low disorder levels, the $T_{MI}$ decreases as predicted, as the maximum of $\alpha_\rho$ decreases. However, for high point defects disorder (e.g. 0.206), $\alpha_\rho$ increases while $T_{MI}$ continue to decrease. This indicates that $\alpha_\rho$ is probably not a good universal parameter of the disorder. The un-monotonous behaviour of $\alpha_\rho$ as function of the disorder is probably related to the percolative character of the metal-insulator transition. There is an important change in the system resistivity when conducting channels are formed, given place to a more abrupt increase in the slope of the resistivity versus temperature curve. More measurements are needed in order to clarify this issue.

At low temperatures, the resistivity is described with a variable range hopping (VRH) model for electronic transport.. (Figure 5). The VRH conductivity for 3D systems is given by [33]:

$$Ln(\sigma) = \ln(\sigma_0) - \left(\frac{T_0}{T}\right)^{1/4} \quad \text{(Ec. 1)}.$$

where $T_0$ is the distance between conducting zones in a 4D space, formed given by the real space distance and the difference of energy. VRH has been observed in cation substituted manganites [34], in substrate-induced disordered films [9] and for $La_{1-x}Pb_xMnO_3$ polycrystalline samples with x = 0.1 and 0.5 [35]. The inset of figure 8 shows the exponential dependence of $T_0$ with increasing point defect disorder. The functional dependence of $T_0$ with the disorder degree, observed in the irradiated films, is the same to the one found for substrate-induced strains. [9]. However, ion irradiation is again probed to be the more efficient way to induce structural disorder, i.e. for the higher irradiation doses the change in the distance between conducting zones is around 100 times bigger than those measured in the thinner films where the substrate-induced strains are more important. CAFM images show that as the irradiation dose increases the distance between conducting regions increases, increasing the localization and the VRH hopping contribution. The increase of the VRH contribution with increasing radiation was also observed for high energy $Li^{3+}$ ion irradiation [36].

Figure 6 presents the magnetoresistance (MR=[ρ(H)-ρ(H=0)]/ρ(H=0)) curves at different temperatures for the sample with 0.206 dpa. Typically the irradiated samples present the same behaviour as pristine manganite films. Remarkably the samples present no sign of low field magneto-resistance (LFMR) as one may expect from an inhomogeneous system. The appearance of the LFMR for irradiated systems has been ascribed to oxygen loss and inhomogeneous defects depth distribution [18]. The absence of LFMR confirms the fact that the irradiation of the samples is macroscopically homogeneous. The MR curves measured at 4 K and at the same field-sweep rate broadens as disorder increases. This is a signature of magnetic relaxation [37] and can be assigned to disorder-induced magnetic frustration or increasing competition between the magnetic interactions in the samples.

Figure 7 presents the MR(H=5T) as function of temperature for manganite films with increasing point defect disorder. As expected [11,14] the temperature of the MR peak decreases for increasing irradiation dose, following the metal-insulation transition. Figure 7 shows that the MR peak broadens for increasing disorder. If the irradiation damage is too great (dpa>0.13) the MR starts to decrease. Surprisingly, even the sample with higher disorder and insulating behaviour showed important MR values. This effect could also be related to the inhomogeneous nature of the sample. For increasing magnetic fields the resistance of the small ferromagnetic and conducting regions decreases. Even if these "islands" don't percolate, the current follows the path of less resistance and flows through these regions, decreasing the total resistance. In this way, the resistivity change of these "islands" may be sensed macroscopically.

The transport properties of irradiated manganite films are understood by taking into account the inhomogeneous nature of the samples. The metal-insulator transition is described by a percolative model where the $T_{MI}$ depends on the distance between conducting regions. As disorder increases this distance increases lowering the observed $T_{MI}$. Increasing the density of point defects increases the mobility edges in the density of states, i.e. the energy required to delocalize the carriers increases. Finally we have found that for low temperatures the main conduction mechanism is variable range hopping with increasing localization as disorder increases.

**III.3 Magnetization properties.**

Figure 8 shows the remnant magnetization as function of temperature for the different irradiated samples. All the samples are ferromagnetic in spite of the fact that the total remnant magnetization decreases as the number of dpa increases. It can be seen that the saturation magnetization decreases exponentially with increasing disorder, while the Curie temperature, $T_C$, remains almost unchanged across the samples series (Figure 9). A slight Tc increase for higher irradiation doses can be ascribed to a strain relaxation through the structural disorder. A noticeable broadening of the magnetic transition is observed as the irradiation dose increases as a consequence of the progressive increase of structural disorder in the samples. It is worth mentioning that there is a strong correlation between the saturation magnetization and the total conducting surface of the samples [24]. This result can be explained by associating conducting regions with the samples ferromagnetism. Since there is no change in Tc for increasing irradiation doses, there is an important decoupling between the $T_{MI}$ and $T_C$ and the difference $T_C$-$T_{MI}$ increases with the density of defects. This tendency has been also observed in strained films [9] and oxygen-deficient films [39] and cannot be explained in the frame of the classical double exchange model [38].

Figure 10 shows the magnetization versus magnetic field at 5 K for manganite films with increasing point defect disorder. The remnant magnetization is close to the saturation magnetization for all the samples and the saturation field of the ferromagnetic component of the samples remains low for all the samples (≤ 2000 Oe). On the other hand, the coercive field of the films presents a maximum as function of the point defects disorder (Figure 11). For low density of defects (dpa<0.15) the coercive field increases linearly with increasing disorder. It is reasonable to think that increasing density of defects enhances the magnetic domain walls pinning, increasing the coercive field. However, when disorder greatly reduces the saturation magnetization and probably the magnetic anisotropy of the ferromagnetic regions a decrease of the coercive field is expected, given place to observed behaviour for the coercive field. We have not observed signs of magnetic frustration or paramagnetic behaviour in our samples, as indicated by the low temperature MR curves, even when performing field and zero field cooling magnetic measurements. It is worth noting however, that a correct quantification of the paramagnetic phase or magnetic frustration is difficult due to the small magnitude of the expected magnetic signals and the substrate contribution. Transport and magnetic relaxation measurements are in progress to address this issue.

Summarizing, the magnetic properties of ion irradiated manganite films can be interpreted considering the inhomogeneous nature of the samples. Ion irradiation creates regions probably frustrated or non-magnetic, surrounding less disordered ferromagnetic regions. As disorder increases the number of the ferromagnetic regions decrease decreasing the saturation magnetization of the system. Strangely, point defect disorder seems to leave the magnetic transition of these regions nearly unchanged.

**IV. Conclusions.**

We have studied the influence of point defect disorder in the transport and magnetic properties of manganite films. Inhomogeneities in the creation of vacancy-interstitial pairs by ion irradiation or changes of the local Mn3+/4+ ratio to reduce lattice strains of the irradiated areas induce electronic inhomogeneities in the films. The transport and magnetic properties of these systems can be interpreted in the context of this scenario. The metal-insulator transition can be described by a percolative model, in which disorder increases the distance between conducting regions, lowering the observed $T_{MI}$. Point defect disorder increases localization of the carriers due to increased disorder and locally enhanced strain field. Increasing vacancy-interstitial disorder also decreases the system magnetization but doesn't seem to change the magnetic transition temperature.

**V. Acknowledgements.**

The authors acknowledge K. Bouzehouane and S. Fusil for the formation received in CAFM measurements. The author would also like to thank R. Benavidez, J.C. Perez, Micra and Veeco crew for extraordinary technical support. This work was partially supported by the ANPCYT (PICT 05-33304, PICT 06- 2092). M. S, N. H E.K. and L.B.S are members of CONICET, Argentina.

**References.**

**Figure 1 :** Total normalized depth distribution of defects calculated by Monte Carlo simulations using the TRIM software (see text). The normalized depth distribution of defects for each unit cell ion specie is also shown.

**Figure 2:** Topographic (left) and CAFM (right) 5 μm x 5 μm images of irradiated LSMO films for different irradiation doses: (a) as-grown , (b) $3.5 \times 10^{14}$ ion/cm$^2$, (c) $7.5 \times 10^{14}$ ion/cm$^2$ and (d) $10 \times 10^{14}$ ion/cm$^2$. Dark areas represent insulating regions and light areas correspond to conducting ones.

**Figure 3:** Zero field resistivity (logarithmic scale) versus temperature (reciprocal scale) for increasing point defect disorder. The arrows indicate the metal-insulator transition.

**Figure 4:** Energy gap,Δ, (left) and $T_{MI}$ (right) vs. dpa of LSMO films. The lines are the fits of the data with linear regressions.

**Figure 5:** Low temperature resistivity as function of temperature for increasing structural disorder. The left insert show an amplification of the box in the main figure. The right inset show the hopping distance as function of the ion irradiated induced point defect disorder. The line is a fitting of the data showing an exponential growing behaviour.

**Figure 6:** Typical magnetoresitance curves at different temperatures for ion irradiated manganite films. The curves corresponding to the sample with a dpa value of 0.206 are shown.

**Figure 7:** Total magnetoresistance at 5 T as function of temperature for ion irradiated films with increasing structural disorder. Lines are guides for the eyes.

**Figure 8:** Magnetization versus temperature for manganite films with increasing point defect disorder (dpa: 0, 0.03; 0.08; 0.15; 0.206; 0.3).

**Figure 9**: Magnetization and Tc (inset) as function of the irradiated manganite films dpa. The line is an exponential fit of the experimental data.

**Figure 10:** Magnetization versus magnetic field for increasing ion irradiated induced structural disorder.

**Figure 11:** Coercive field as function of the point defect disorder for ion irradiated manganite films. Lines are only guides for the eyes.

Figure 1: Sirena at. al

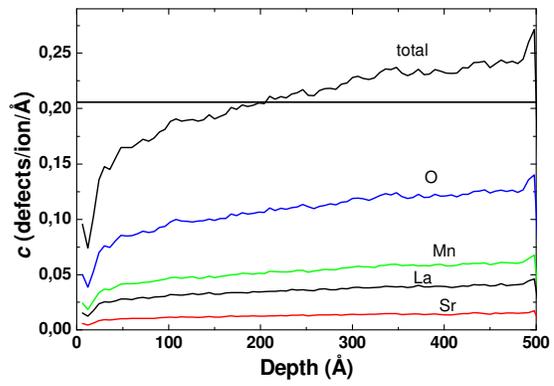

Figure 2: Sirena et. al.

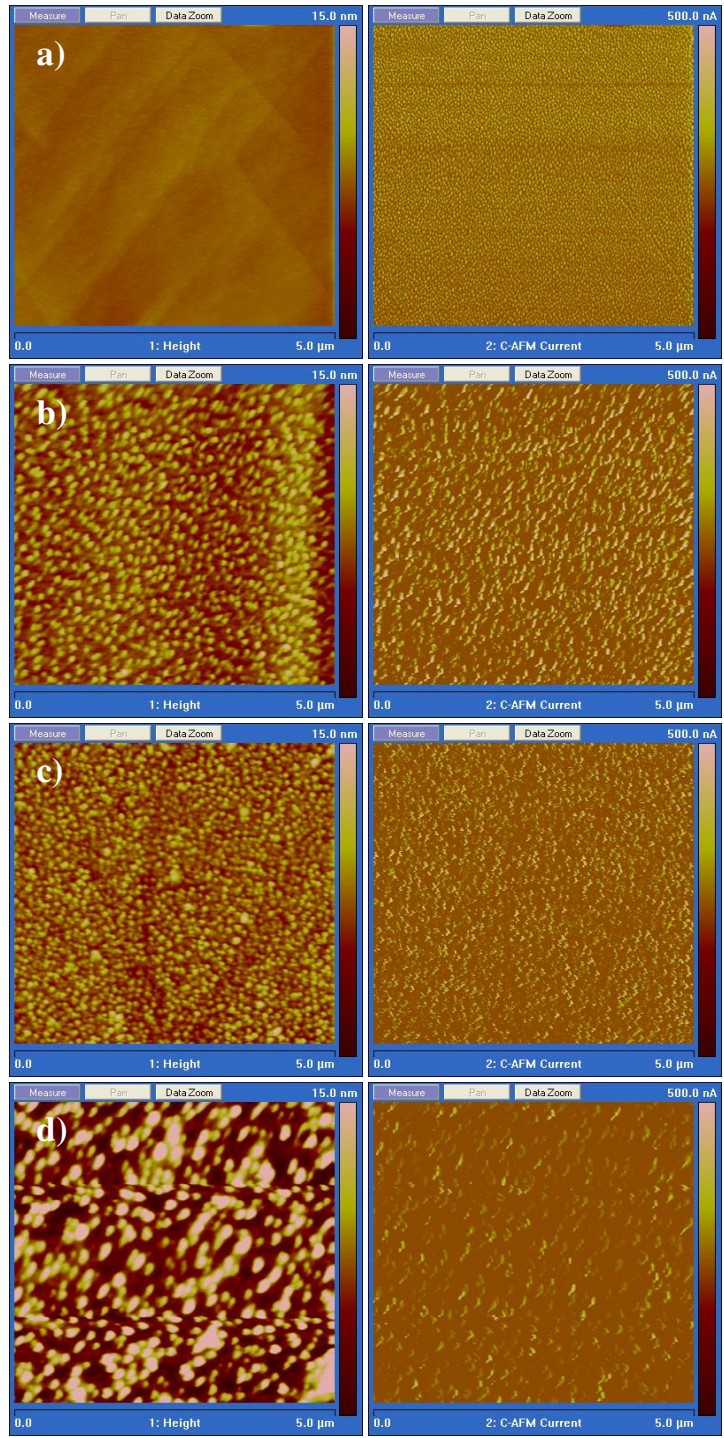

Figure 3: sirena et. al.

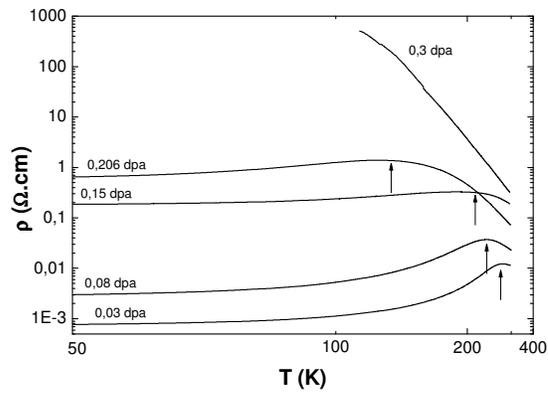

Figure 4 : sirena et. al.

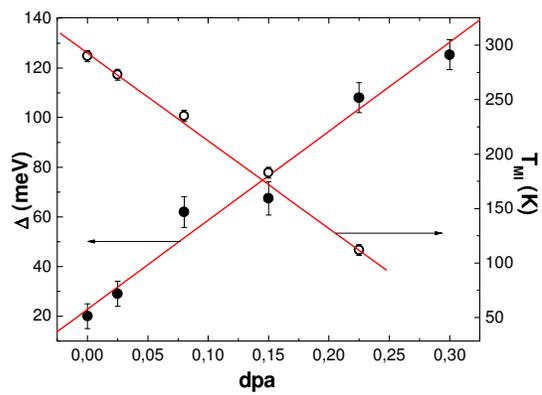

Figure 5 : Sirena et. al

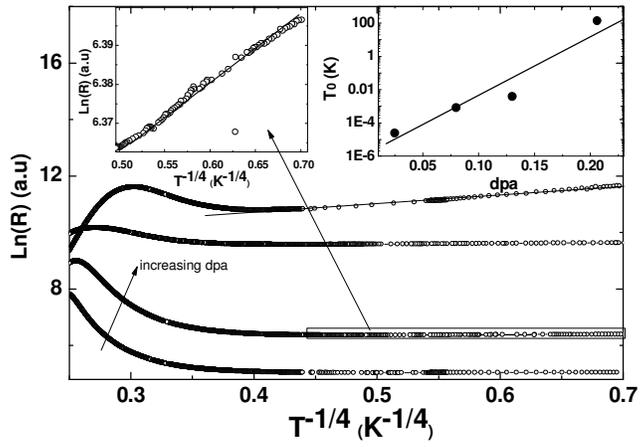

Figure 6 : Sirena et al

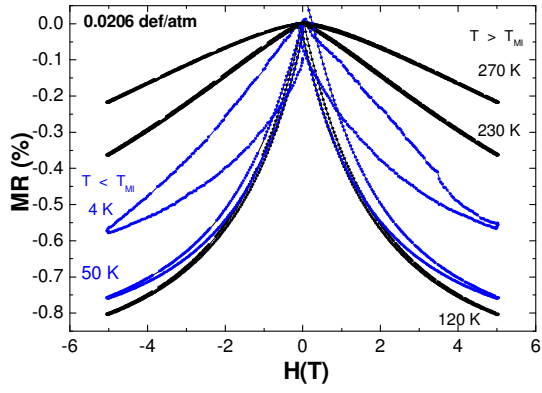

Figure 7 : Sirena et. al

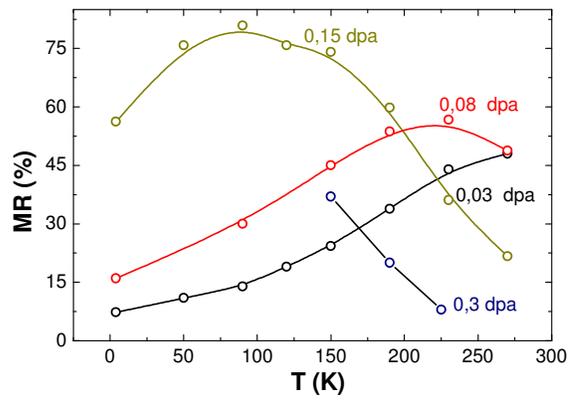

Figure 8 : Sirena et. al ;

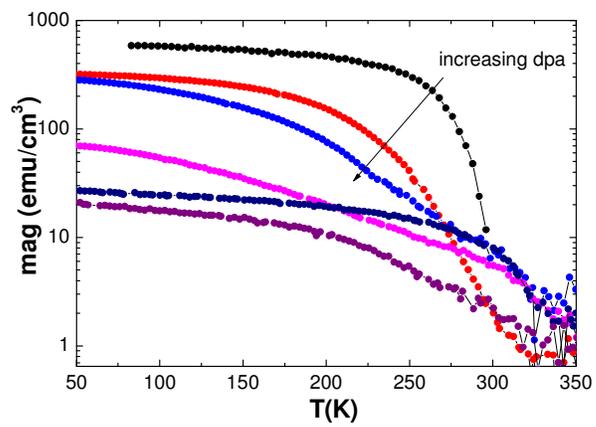

Figure 9: Sirena et. al

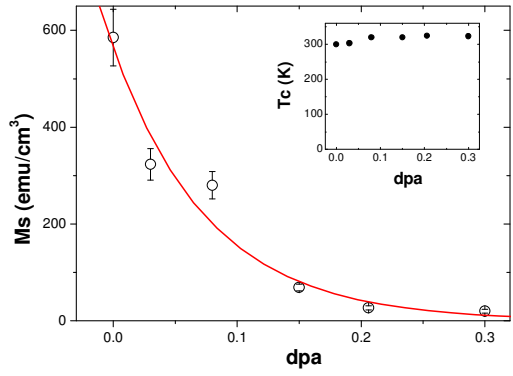

Figure 10: Sirena et. al.

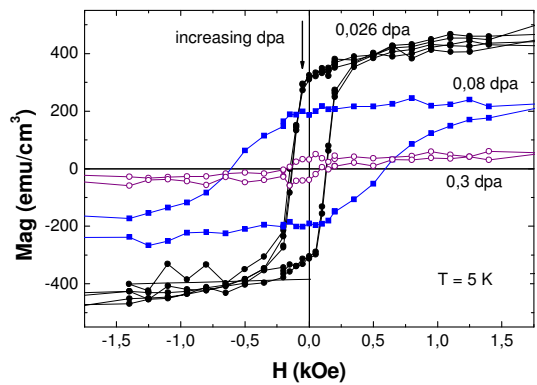

Figure 11: Sirena et. al.

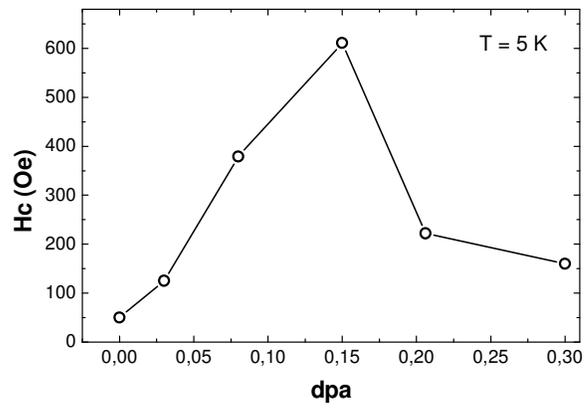